\documentclass{article}

\makeatletter

\@addtoreset{equation}{section}
\makeatother
\newtheorem{theo}{Theorem}[section]
\newtheorem{rema}[theo]{Remark}

\newtheorem{propo}[theo]{Proposition}
\newtheorem{coro}[theo]{Corollary}
\newtheorem{defi}[theo]{Definition}
\newtheorem{ex}[theo]{Example}
\newcommand{\m}{{\cal M}}   
\newcommand{\mo}{{\cal M}_0}  
\newcommand{\cat}{{\rm cat}}    
\newcommand{\be}{\begin{equation}}
\newcommand{\ee}{\end{equation}}
\newcommand{\proof}{\noindent{\bf Proof.\ }}
\newcommand{\cvd}{{\rule{0.5em}{0.5em}}\medskip}

\font\ddpp=msbm10 
\def\R{\hbox{\ddpp R}}    

\def\N{\hbox{\ddpp N}}    

\title{{\bf On general Plane Fronted Waves. Geodesics}}

\author{A.M. CANDELA\thanks{Supported by M.I.U.R. (research funds ex 40\% and
60\%).} ,
J.L. FLORES$^\dagger$,
M. S\'ANCHEZ \thanks{ Partially supported by a MCyT-FEDER Grant BFM2001-2871-C04-01.}\\
\\
$^*$ {\small Dipartimento Interuniversitario di Matematica,}\\
{\small Universit\`a degli Studi di Bari,}\\
{\small Via E. Orabona 4,
70125 Bari, Italy}\\
$^\dagger$ {\small Departamento de Geometr\'{\i}a y Topolog\'{\i}a,}\\
{\small Facultad de Ciencias, Universidad de Granada,}\\
{\small Avenida Fuentenueva s/n, 18071 Granada, Spain.}}

\date{}

\begin{document}
\textwidth=140mm 
\textheight=200mm 
\parindent=5mm
\maketitle

\begin{quote}

\noindent {\small \bf Abstract.} {\small A general class of
Lorentzian metrics, $\mo \times \R^2$,
$\langle\cdot,\cdot\rangle_z = \langle\cdot,\cdot\rangle_x + 2\
du\ dv + H(x,u)\ du^2$, with $(\mo, \langle\cdot,\cdot\rangle_x) $
any Riemannian mani\-fold, is introduced in order to generalize
classical exact plane fronted waves. Here, we start a systematic
study of their main geodesic properties: geodesic completeness,
geodesic  connectedness and multiplicity or causal character of
connecting geodesics. These results are independent of the
possibility of a full integration of geodesic equations.
Variational and geometrical techniques are applied systematically.
In particular, we prove that the asymptotic behavior of $H(x,u)$
with $x$ at infinity determines many properties of geodesics.
Essentially, a subquadratic growth of $H$ ensures geodesic
completeness and connectedness, while the critical situation
appears when $H(x,u)$ behaves in some direction as $|x|^2$, as in
the classical model of exact
gravitational waves.}
\end{quote}

\section{Introduction}

Aim of this paper is to study some global properties of a family of Lorentzian manifolds
which  model plane fronted waves and, in particular, gravitational waves.

As explained in the classical book by Misner, Thorne and Wheeler \cite{MTW},
a gravitational wave represents ripples in the shape of spacetime which
propagate across spacetime, as water waves are small ripples in the shape
of the ocean's surface propagating across the ocean. The source of a gravitational wave
is the motion of massive particles; in order to be detectable, very massive objects under
violent dynamics must be involved (binary stars, supernovae,
gravitational collapses of stars...). With more generality, plane fronted waves may also
take into account the propagation of non--gravitational effects such as electromagnetism.
 {\em Exact plane fronted wave} solutions
arise as a class of highly idealized standard models, as a
compromise between reality and simplicity. Concretely, this model
is a Lorentzian manifold $(\R^4,ds^2)$ endowed with the  metric
\be \label{planefrontedmetric} d s^2 = dx_1^2 + dx_2^2 + 2\ du\ dv
+ H(x_1, x_2,u)\ du^2 , \ee $(x_1, x_2,v,u) \in \R^4$, where $H :
\R^3 \to \R$ is a non--null smooth function. The scalar curvature
of these spacetimes is always zero, but they are Ricci flat if and
only if \be \label{elapla} \Delta_x H(x,u) \equiv 0, \ee where
$\Delta_x H$ denotes the Laplacian of $H$ with respect to $x=(x_1,
x_2)$ (gravitational $pp$-wave). When one considers exact
electromagnetic or gravitational waves, $H(\cdot, u)$ is assumed
to be a quadratic form on $\R^2$, with the additional assumption
(\ref{elapla}) in the gravitational case. Thus, an {\sl exact
(plane fronted) gravitational wave} is the particular spacetime
obtained when the coefficient $H$ in (\ref{planefrontedmetric})
has the special form \be \label{gravitational} H(x_1, x_2,u) =
f(u)(x_{1}^{2} - x_{2}^{2}) + 2 g(u)\ x_{1}x_{2} \ee for some $f$,
$g \in C^2(\R,\R)$, $f^2 + g^2 \not\equiv 0$ (an additional
condition commonly used in this case is $\int^{+\infty}_{-\infty}
\root 4\of{f^2(u) + g^2(u)}du < +\infty$). In particular, an exact
gravitational wave is a {\sl sandwich wave} if both $f$ and $g$
have compact support, while it is a {\sl polarized wave} if $g
\equiv 0$.

Historically, the study of gravitational waves goes back to
Einstein (cf. \cite{ER}) but the standard exact model was already
introduced by Brinkmann in order to determine Einstein spaces
which can be improperly mapped conformally on some Einstein one
(see \cite{Br}). Afterwards, they have been widely studied by many
authors (see, for example, the summary in \cite{Yu}). From the
experimental viewpoint, the detection of gravitational waves is
now one of the most exciting challenges\footnote{Hulse and Taylor
were awarded the Nobel Prize in 1993 for the seminal discovery in
the seventies of indirect evidences of their existence; there is
now a world wide effort -projects as LISA, GEO600, VIRGO, LIGO or
TAMA300- for direct detection.}.

In a series of articles, Ehrlich and Emch
studied systematically global properties of the exact model (cf. \cite{EE,EE2a,EE3}).
Especially, they studied  geodesics of gravitational waves by means of the
symmetries of the model, which allow an almost full integration of the geodesic
equations. Summing up, important goals have been the characterization of
properties such as causality, geodesic completeness, (non) geodesic
connectedness or astigmatic conjugacy (all of them  explained pedagogically in the book \cite{BEE}).

Even though Ehrlich and Emch's study is very complete and exhaustive, the dependence of their
results on the high degree of symmetry of the standard exact model must come in mind. Thus, in the
present paper we study the following generalization of the classical model.

\begin{defi}
\label{manifold}
A semi--Riemannian manifold $(\m,\langle\cdot,\cdot\rangle_z)$
is a plane fronted wave, briefly PFW, if there exists a connected
$n$--dimensional Riemannian mani\-fold
$(\mo,\langle\cdot,\cdot\rangle_x)$
such that it is $\m = \mo \times \R^2$ and
\be
\label{metric}
\langle\cdot,\cdot\rangle_z = \langle\cdot,\cdot\rangle_x
+ 2\ du\ dv + H(x,u)\ du^2 ,
\ee
where $x \in \mo$, the variables $(v,u)$ are the natural coordinates
of $\R^2$ and the smooth scalar field
$H : \mo \times \R \to \R$ is such that $H \not\equiv 0.$
\end{defi}
Let us remark that if $\mo = \R^2$ is the standard 2--dimensional
Euclidean space, the metric (\ref{metric}) reduces to
(\ref{planefrontedmetric}) so, throughout all this paper, {\sl
exact PFW} will mean $\R^4$ endowed with
(\ref{planefrontedmetric}), while for an {\sl exact  gravitational
wave} also (\ref{gravitational}) holds. Furthermore, for
simplicity, in what follows the subscript of the metrics
$\langle\cdot,\cdot\rangle_{z}$, $\langle\cdot,\cdot\rangle_{x}$
($z=(x,v,u)$) will be dropped without possibility of confusion.

Definition \ref{manifold} is a convenient generalization
under both the physical and the mathematical viewpoint.
Recall that, from the physical viewpoint, the existence
of many symmetries must be regarded only as a first
approach. But these symmetries cannot be expected to
happen in an exact way; therefore,  physical results
must be  independent  of them in some reasonable sense.
Our generalization retains the characteristic structure of
an exact plane fronted wave, but it drops additional symmetries.
Notice also that many authors have introduced modifications in
the exact model in order to describe different situations
such as, for example, colliding waves, or Schwarzschild or
de Sitter background (see, e.g., \cite{BH,BFI,Bo,CV,Gr,Ho,Kr,Yu,Za});
moreover, the unrestricted topology of $\mo$ may be useful not only
for such models but also for quantization (see, e.g., \cite{Ez}). From the purely mathematical viewpoint, recall that now fundamental
equations, as geodesic equations, cannot be integrated explicitly.
Thus, one can see exactly the different mathematical tools needed
for the different mathematical problems, as well as the exact
relations among these results.

\smallskip

\noindent This paper and a forthcoming one (see \cite{FS})
are devoted to study systematically some geometrical properties of PFWs.
In the present article, general properties and, especially, those
ones concerning in geodesics are analyzed. It
is organised as follows.

In Section \ref{s2}, the Levi-Civita connection is determined.
The scalar curvature is equal to that one of the Riemannian part
and the assumptions for being Ricci flat or
for satisfying the timelike convergence condition are given
(see Propositions \ref{p2.1}, \ref{p2.2} and Remark \ref{r2.3}).

In Section \ref{s3}, a preliminary study of geodesics is carried
out. Geodesics in a PFW are related to trajectories of a particle
on the Riemannian manifold $(\mo, \langle \cdot , \cdot \rangle)$
under a potential $V_\Delta$ depending on time, where $V_\Delta$
is essentially equal to $-H$ (see Proposition \ref{pr3.1}). As a
consequence, geodesic completeness is related to the completeness
of trajectories for this potential (see Theorem \ref{th3.2}). A
natural condition for geodesic completeness is then {\em positive
completeness} which holds, in particular, if $H(x,u)$ does not
increase superquadratically with $x$ in some direction and its
growth with respect to $u$ is bounded (see Proposition
\ref{libro}, Corollary \ref{co3.3}). Remarkably, for an exact
gravitational wave the behavior of the function $H$ with respect
to the variable $u$ may lie out of these hypotheses. Even though
its completeness can be obtained in a straightforward way (see
Proposition \ref{co3.4}), we emphasize that such a result relies
on the very especial form of $H$.

Section \ref{s40} is devoted to connectedness by geodesics.
As a difference with previous references on exact PFWs, the
impossibility to integrate geodesic equations force us to use
results from global variational methods and Ljusternik-Schnirelman theory.

In Subsection \ref{s4} geodesics connecting two fixed points $z_0, z_1 \in \m$
are related to the existence of critical points for a Lagrangian functional
${\cal J}_{\Delta}$ depending only on the Riemannian part
and we prove (see Theorem \ref{pisi}, Corollary \ref{corocon}):
(i) the existence of at least one connecting geodesic from $z_0$ to $z_1$
(i.e., geodesic connectedness) whenever $H$ does not become negative and
decreasing with $x$ quadratically or faster  (i.e., essentially,
$H(x,u) \geq -R_0(u)|x|^{2-\epsilon}$ for some $\epsilon>0$) and (ii) under the previous assumptions,
the existence of infinitely many spacelike connecting geodesics, if the
topology of $\mo$ is not homotopically trivial.
Let us point out that exact gravitational waves are examples of PFWs with a
quadratic growth of $H$ in some directions, {\em which are not geodesically connected}.
Nevertheless, even in this case, Theorem \ref{pisi}
gives an estimate of which points can be connected
by a geodesic, which is shown to be optimal (see Example \ref{esempio}).

In Subsection \ref{s5} connectedness by causal geodesics is studied.
Recall that a classical theorem by Avez and Seifert asserts:
{\em in a globally hyperbolic spacetime, any pair of causally
related points (i.e., a pair of points which can be joined by
a causal curve) can be joined by a causal length-maximizing geodesic}
(see \cite{Av,Se}).
It is known that this conclusion does not hold for an exact gravitational
wave. Nevertheless, in Theorem \ref{prrr} we give an optimal estimate for
the points where the conclusion holds, valid in general PFWs, while
in Theorem \ref{prrr2} and Corollary \ref{prrr1} some multiplicity results
are stated.

In Subsection \ref{s6} all the results in the previous two subsections
are applied to the special case of exact gravitational waves, giving an
accurate estimate about which points can be geodesically connected with
others, the causal character of the connecting geodesics and its possible multiplicity
(see Corollary \ref{coro0}, Proposition \ref{c1} and Remark \ref{rema1}).

\smallskip

\noindent It is worth pointing out that all these results on geodesics depend only
on the qualitative behavior of $H$ at infinity. Thus, they are independent
of a property such as the focusing effect of null geodesics, a folk
characteristic property of exact gravitational waves since
Penrose's article \cite{Pe}. In fact, it is easy to find examples
of PFWs (Ricci flat --vacuum-- or satisfying the timelike convergence
condition) whose geodesics satisfy only some selected properties
(completeness, geodesic connectedness).
Further discussions in \cite{FS} will show how, although focusing
effect is also related to properties as (the lack of) global
hyperbolicity in the exact case, it is rather independent in the generic
non--exact case.


\section{Christoffel symbols and curvature} \label{s2} From now on, manifolds are $C^3$ while functions are $C^2$ as a simplification,
even though we need only $C^1$ for many purposes.
Furthermore, we will say that a tangent vector $w$
is lightlike if $\langle w, w\rangle =0$ and $w\neq 0$; while
$w$ is causal if it is either lightlike or timelike (0 is spacelike).

Now, let ${\cal M}={\cal M}_{0}\times\R^{2}$ be a PFW
equipped with the metric (\ref{metric}).
We can fix a time orientation on it such that the lightlike vector field
$\partial_v$  is past directed; thus, the lightlike vector field
$\partial_u -\frac{1}{2} H\partial_v$ will be future directed.
It is easy to check that $\partial_v$ is also a parallel
vector field and $\partial_v = \nabla u$,
where $u$ is the projection
\[
(x,v,u) \in \mo\times \R^2 \longmapsto u \in \R.
\]
Thus, for any
future directed causal curve $z(s)=(x(s),v(s),u(s))$,
there results
\be
\label{udot}
\dot u(s) = \langle \dot z(s), \partial_v\rangle \geq 0,
\ee
and the inequality is strict if $z(s)$ is timelike
(the assumed background for this paper can be found
in well--known books as \cite{BEE,HE,ON}).

Fix some local coordinates $x^1, \dots, x^n$ with respect to the Riemannian part $\mo$, as well as
 $(v,u)\in\R^2$. A direct computation shows that the  non--necessarily null
Christoffel's symbols are
\begin{eqnarray*}
&&\Gamma^k_{i j} = \Gamma^{k(R)}_{i j} \qquad \mbox{for all $k, i, j \in \{1, \dots, n\}$,}\\
&&\Gamma^v_{u j} = \Gamma^v_{j u} =\ {1\over 2}\ {\partial H\over \partial x^j} (x,u)
\qquad \mbox{for all $j \in \{1, \dots, n\}$,}\\
&&\Gamma^v_{u u} = \ {1\over 2}\ {\partial H\over \partial u} (x,u),\\
&&\Gamma^k_{u u} = - \ {1\over 2}\ \sum_{m=1}^n g_{(R)}^{km}\ {\partial H\over \partial x^m} (x,u)
\qquad \mbox{for all $k \in \{1, \dots, n\}$,}\\
\end{eqnarray*}
where $(g_{(R)}^{ij})_{ij}$ is the inverse of the matrix associated to the Riemannian
metric on $\mo$ and
$\Gamma^{k(R)}_{i j}$ are its Christoffel's symbols if
$k, i, j \in \{1, \dots, n\}$.
Thus, the only non--null components of the Ricci curvature of
the metric are
\begin{eqnarray*}
&&R_{i j} = R^{(R)}_{i j} \qquad \mbox{for all $i, j \in \{1, \dots, n\}$,} \\
&&R_{u u} =  - \ {1\over 2}\ \left(\sum_{k,l=1}^n {\partial\over \partial x^k}
\big(g_{(R)}^{k l}\ {\partial H\over \partial x^l} (x,u)\big)\ +\
\sum_{k,a, l=1}^n g_{(R)}^{a l}\ {\partial H\over \partial x^l} (x,u)\Gamma^{k(R)}_{k a}\right)
\\
&&\quad \quad =-\frac{1}{2} \Delta_{x} H(x,u),
\end{eqnarray*}
where $R^{(R)}_{i j}$ and $\Delta_{x}$ are the components of the Ricci
curvature and the Laplacian, respectively, associated to the Riemannian metric on $\mo$.
Thus, the (local form of the) Ricci curvature is
\begin{equation}\label{rtr}
{\rm Ric} = \sum_{i,j=1}^n R^{(R)}_{i j} d x^i \otimes d x^j -\frac{1}{2}\Delta_{x}H du \otimes du .
\end{equation} From (\ref{rtr}) it is easy to check the following two propositions:

\begin{propo}
\label{p2.1}
In a PFW:
\begin{enumerate}
\item[(i)] the scalar curvature at each $(x,v,u)$ is equal to the
scalar curvature of the Riemannian part
$(\mo, \langle \cdot, \cdot \rangle)$ at $x$;

\item[(ii)] the  Ricci tensor Ric is null if and only if the Riemannian
Ricci tensor Ric$^{(R)}$ is null and $\Delta_{x}H \equiv 0$.
\end{enumerate}
\end{propo}

\begin{propo}
\label{p2.2}
A PFW satisfies the timelike convergence condition (i.e., for all
timelike vector $\zeta$, Ric$(\zeta,\zeta)\geq 0$) if and only if
for all $(x,u) \in \mo\times\R$ and
$w\in T_{x}{\cal M}_{0}$, $w\neq 0$, it is
\begin{eqnarray}
&&\Delta_{x}H(x,u)\leq 0\quad \mbox{and}\label{ii}\\
&&Ric^{(R)}(w,w)\geq 0. \label{iii}
\end{eqnarray}
\end{propo}

\begin{rema}
\label{r2.3}
{\em It is well--known that if a function $f$ on a connected Riemannian
manifold satisfies $\Delta f \leq 0$ (or $\geq 0$) and
$\nabla f \equiv 0$  out of a compact subset, then $f$ is constant.
Thus, if for each fixed $u$, $\nabla_xH(\cdot,u)$ is zero out of a
compact subset (in particular, if $\mo$ is compact), then
condition (\ref{ii}) implies $H(x,u)\equiv H(u)$.

If $Ric^{(R)}(w,w)\geq\epsilon >0$ for all unit $w$ and the
Riemannian metric on ${\cal M}_{0}$ is complete, then $\mo$ is
compact by Bonnet--Myers theorem; thus,  condition (\ref{ii})
would imply that $H$ is independent of $x$ (notice that
$Ric^{(R)}(w,w)>0$ for all $w$ is possible in a complete
non-compact manifold; for example, this happens in a paraboloid).

Recall that, when $H(x,u)\equiv H(u)$, the corresponding PFW is the product of
the Riemannian part $\mo$ by $\R^2$ endowed
with the metric $2dudv + H(u) du^2$ (such a bidimensional
metric is flat, but simple natural extensions has its own interest, cf. \cite{RS,Sa-trans}).}
\end{rema}

\section{General behavior of geodesics.\\ Completeness} \label{s3}

Aim of this section is analizing the behavior of geodesics
in a PFW and, in particular, studying their completeness,
pointing out some sufficient conditions to geodesic completeness
of the manifold.

\begin{propo}
\label{pr3.1}
Let $z:\ ]a,b[\ \rightarrow \m$,
$z(s)= (x(s), v(s), u(s))$ ($\ ]a,b[\ \subseteq \R$),
be a curve on $\m$ with constant energy $\langle \dot z(s), \dot z(s)\rangle = E_z$
for all $s \in \ ]a,b[$ .
Assume $0 \in\ ]a,b[\ $.
Then, $z$ is a geodesic on $\m$ if and only if the three following conditions hold:
\begin{enumerate}
\item[$(a)$]
$u = u(s)$ is affine, i.e.,
$u(s) = u_0 + s \Delta u$ for all $s \in \ ]a,b[$,
where $u_0 = u(0)$, $\Delta u = \dot u(0)$;

\item[$(b)$]
$x = x(s)$ is a solution of
\begin{equation}
\label{RiemannianEq}
D_s\dot x = - \nabla_x V_{\Delta}(x(s),s) \quad
\mbox{for all $s \in \ ]a,b[$,}
\end{equation}
where
\begin{equation} \label{eV}
V_{\Delta}(x,s) = -\ \frac{(\Delta u)^2}{2}\ H(x, u_0 + s \Delta u);
\end{equation}

\item[$(c)$]
if $\Delta u = 0$ then $v = v(s)$ is affine, i.e.,
$v(s) = v_0 + s \Delta v$ for all $s \in\ ]a,b[$, with $v_0 = v(0)$, $\Delta v = \dot v(0)$;
otherwise, for all $s \in \ ]a,b[$ it is
\[
v(s) = v_0 + \frac{1}{2 \Delta u} \int_0^s \left( E_z - \langle \dot x(\sigma), \dot x(\sigma)\rangle +
2 V_{\Delta}(x(\sigma), \sigma)\right) d\sigma.
\]
\end{enumerate}
\end{propo}

\proof
Fix local coordinates $(x^1, \dots, x^n,v,u)$, as in Section \ref{s2}, and consider
Christoffel's symbols of the metric (\ref{metric}); the geodesic equations become
\begin{eqnarray}
&&
\ddot x^i + \sum_{j,k=1}^n \Gamma^{i(R)}_{j k} \dot x^j \dot x^k + \Gamma^{i}_{u u} \dot u^2 = 0
\qquad \mbox{for all $i \in \{1, \dots, n\}$,}\label{primo}\\
&&\ddot v + 
2\ \sum_{j=1}^n \Gamma^{v}_{j u} \dot x^j \dot u + \Gamma^{v}_{u u} \dot u^2 = 0,
\label{secondo}\\
&& \ddot u = 0.\label{terzo}
\end{eqnarray}
Thus, $(a)$ and $(b)$ follows from (\ref{primo}), (\ref{terzo}). For $(c)$,
the expression of the energy $E_z$ yields
\begin{equation}
\label{quarto}
2\dot v \dot u = E_z - \langle \dot x, \dot x\rangle - H(x,u) \dot u^2.
\end{equation}
So, if $\Delta u =0$ by $(a)$ it is $\dot u\equiv 0$,
so, use (\ref{secondo}); otherwise, use (\ref{quarto}).
\cvd

Therefore, in order to investigate the properties of geodesics
in a PFW, it is enough studying the behavior of the Riemannian trajectories
under a suitable potential $V\equiv V_{\Delta}$. In particular, this happens
for geodesic completeness, where we can take just $V=-H$ (or, for convenience, $V=-H/2$).

\begin{theo}
\label{th3.2}
A PFW is geodesically complete if and only if $(\mo,\langle\cdot,\cdot\rangle)$
 is a complete
Riemannian manifold and the trajectories of
\begin{equation}
\label{ev1}
D_s\dot x = \ \frac 1 2\ \nabla_x H(x,s)
\end{equation}
are complete, i.e., each of them can be extended so as to be defined on all $\R$.
\end{theo}

\proof
The implication to the right is obvious by Proposition \ref{pr3.1}, as
(\ref{ev1}) is equivalent to (\ref{RiemannianEq}) with $u_0 = 0$ and $\Delta u = 1$,
while each Riemannian geodesic $x = x(s)$ in $\mo$ defines
a trivial geodesic $z = (x,0,0)$ in $\m$.

On the contrary, let $x = x(s)$ be a solution of (\ref{RiemannianEq}).
Then, either $\Delta u = 0$ and $x$ is a Riemannian geodesic in the complete manifold
$\mo$, or $\Delta u \ne 0$ and $y(\sigma) = x((\sigma - u_0)/ \Delta u)$
solves (\ref{ev1}). In both these cases,
$x$ can be extended so as to be defined on all $\R$.
\cvd

The completeness of the trajectories satisfying (\ref{ev1})
has been studied by several authors (see, e.g., \cite{Eb,Go,WM}).
Frequently, they exploit the idea that the velocities of these trajectories
are integral curves of a vector field $X$
(the Lagrangian vector field) on the tangent manifold $T\mo$.
In general, if an integral curve $c = c(s)$ of any vector field is defined on an interval
$[0,b[$, $b < +\infty$, and there exists a sequence $s_n \rightarrow b$ such that $\{c(s_n)\}_n$ converges, then $c$ is extendible as an integral curve beyond $b$ (symmetrically, if $c$ is defined
in $\ ]a,0]$, $a > -\infty$).
Thus, one has just to ensure that the integral curves of $X$
restricted to a bounded interval lie in a compact subset of $T\mo$.

So, if, for example, the Riemannian part $(\mo, \langle \cdot , \cdot \rangle)$ is complete
and the coefficient $H(x,u)$ in (\ref{metric}) is autonomous
(i.e., independent of $u$), a natural condition for completeness of solutions
of (\ref{ev1}) is obtained assuming that $H$ is controlled at infinity (in a suitable way)
by a {\em positively complete} function $U_0$, i.e.,
a nonincreasing $C^2$ function $U_0 : \R_+ \to \R$ ($\R_+ = [0,+\infty[$)
such that
\[
\int_0^{+\infty} \frac {dt}{\sqrt{\alpha - U_0(t)}}\ = +\infty,
\]
for some (and thus any) $\alpha > U_0(0) = \sup U_0(\R_+)$. More precisely, the following result can be stated (see \cite{WM} or also \cite[Theorem 3.7.15]{AM}):

\begin{propo}
\label{libro}
Let $\m = \mo \times \R^2$ be a PFW such that
$(\mo, \langle \cdot , \cdot \rangle)$ is complete
and $H$ is autonomous, i.e., $H(x,u) \equiv H(x)$.
If there exist $r > 0$, $\bar x \in \mo$ and a positively
complete function $U_0 : \R_+ \to \R$ such that
\[
H(x) \le - U_0(d(x,\bar x)) \qquad \mbox{for all $x \in \mo$ such that $d(x,\bar x) \ge r$,}
\]
then all the trajectories which satisfy
(\ref{ev1}) are complete.
\end{propo}
Let us point out that a simple example of positively complete function
is $U_0(t)= - R_0 t^p$ with $R_0 > 0$ and $0 \leq p \leq 2$. So,
Theorem \ref{th3.2} and Proposition \ref{libro}
imply:

\begin{coro} \label{co3.3}
Let $\m = \mo \times \R^2$ be a PFW such that $(\mo, \langle \cdot , \cdot \rangle)$ is complete
and $H(x,u) \equiv H(x)$.
If there exist $r > 0$, $\bar x \in \mo$ and $R_0 > 0$
such that
\[
H(x) \le R_0 d^2(x,\bar x) \qquad \mbox{for all $x \in \mo$ such that $d(x,\bar x) \ge r$,}
\]
then $\m$ is geodesically complete.
\end{coro}

As pointed out in \cite{Go}, the results on autonomous potentials
imply also results on non--autonomous potentials
by considering the product manifold $\mo \times \R$;
essentially, one has also to bound the growth of the potential $V(x,u)$ with $u$.
Nevertheless, the completeness of exact gravitational waves does not seem to be covered
by these general results, at least if $f$ and $g$
in (\ref{gravitational}) are arbitrary (of course, the completeness would be straightforward
for a sandwich wave).
Anyway, it is easy to give particular results on geodesic
completeness which cover all gravitational waves. But they
rely in very particular expressions of $H(x,u)$
and would not hold for an arbitrary (even exact) PFW.
For the sake of completeness, we give such a result.

Consider $\mo = \R^n$ with its classical Euclidean metric $\langle \cdot , \cdot \rangle_{0}$.
Assume that the coefficient $H(x,u)$ in (\ref{metric}) is
in the canonical form
for gravitational or electromagnetic waves, i.e., let
\begin{equation}
\label{canonical}
H(x,u) = \langle A(u) x,x\rangle_{0},
\end{equation}
where $A(u)$ is a non--identically vanishing map
from $\R$ to $M^{\rm sym}(n,\R)$, the space of symmetric $n \times n$ real valued
matrices
(if $n = 2$ and $\Delta_x H(x,u) \equiv 0$, then
we have exactly (\ref{gravitational}); compare with  \cite[Remark 2.3]{EE}).
Under this assumption, it is $\nabla_xH(x,u) = 2 A(u) x$ for all $x \in \R^n$, $u \in \R$,
so the equation (\ref{ev1}) becomes
\begin{equation}
\label{qui}
\ddot x(s) = \ A(s) x(s), \qquad s \in \R .
\end{equation}
A classical global existence theorem for linear ODEs implies
that all the solutions of (\ref{qui}) are complete
so the following result can be stated:

\begin{propo}
\label{co3.4}
Consider a PFW such that $(\mo , \langle \cdot , \cdot \rangle)$ is covered by
 Euclidean space $\R^n$, and $H(x,u)$ is in the canonical form
(\ref{canonical}). Then, the PFW is geodesically complete.

In particular, any exact gravitational wave is geodesically complete.
\end{propo}

Finally, fix $u_0 \in \R$ and consider the hypersurface
\[
\Pi_{u_0} = \{(x,v,u) \in \m : u = u_0\}.
\]
Clearly, the restriction of the metric to this hypersurface is degenerate
positive--semidefinite, so it is not a semi--Riemannian submanifold.
Nevertheless, $\Pi_{u_0}$  is totally geodesic in the sense that taken
any tangent vector $ w $ to $\Pi_{u_0}$ the (necessarily non-timelike)
geodesic in the PFW with initial velocity $ w$ remains in $\Pi_{u_0}$.
More precisely, as a straightforward consequence of Proposition \ref{pr3.1}:

\begin{coro}
\label{corocoro}
Fix a vector $w$ tangent to $\mo$, and $v_0, \Delta v \in \R$.
Let $x(s)$ be the geodesic in $(\mo, \langle\cdot,\cdot\rangle)$
with initial velocity $w$.
Then, the curve in $\Pi_{u_0}$, $z(s)=(x(s), v_0 +  s \Delta v, u_0)$, is a geodesic in the PFW.

Furthermore, a curve $\gamma(s)$ in $\Pi_{u_0}$  is lightlike (at
any point) if and only if it is a reparametrization of a geodesic
$z(s)=(x(s), v_0 + s\Delta v, u_0)$ with constant $x(s)$ and
$\Delta v \neq 0$.
\end{coro}

\section{Connection by geodesics} \label{s40}

Geodesic connectedness of spacetimes has been widely studied
under very different techniques (see, e.g., the survey \cite{San}).
In particular, since the seminal articles by Benci, Fortunato and Giannoni,
variational methods have been extensively used in Lorentzian Geometry
for this and other related problems
(see \cite{BF,BFG} or the book \cite{Ma}).
Nevertheless, here our viewpoint is rather different
and relies exclusively in the previous results by the
authors in \cite{CFS}, which are based in standard variational techniques and
Ljusternik-Schnirelman theory
as developed in references like \cite{FH,Pa1}.

\subsection{Geodesic connectedness} \label{s4}

In what follows, put  $I=[0,1]$ (closed interval)
and recall that, in Proposition \ref{pr3.1},
if $1 \in\ ]a,b[\ $ then $\Delta u = u(1) - u(0)$.
As in the case of geodesic completeness,
the problem of existence and multiplicity of connecting
geodesics in a PFW reduces to the existence and multiplicity
of classical solutions of a Riemannian problem.

\begin{propo}
\label{prlema}
For any PFW the two following properties are equivalent:
\begin{enumerate}
\item[$(a)$] geodesic connectedness (i.e., each two of its points can be joined by a geodesic);

\item[$(b)$] the problem
\begin{equation}
\label{pissi}
\left\{
\begin{array}{l}
D_s\dot x (s)= - \nabla_x V_{\Delta}(x(s),s)\quad \mbox{for all $s \in I$}\\
x(0) = x_0 ,\;\; x(1) = x_1,
\end{array}
\right.
\end{equation}
admits a solution for all
$x_0, x_1 \in \mo$, all the possible values $\Delta u \in \R$ and
all the initial points $u_0 =u(0)$,
where $V_{\Delta}(x,s)$ is given in (\ref{eV}).
\end{enumerate}
\end{propo}

\proof
$(a) \Rightarrow (b)$ Recall that if a geodesic connects two given points,
it can be reparametrized so to make its domain equal to $I$; then, use Proposition \ref{pr3.1}.

$(b) \Rightarrow (a)$ Fixed two points $z_0= (x_0, v_0, u_0), z_1 = (x_1, v_1, u_1)$,
the connecting geodesic $z(s)=(x(s), v(s), u_0 + s \Delta u)$, $\Delta u =u_1-u_0$,
is obtained taking $x(s)$ as solution of (\ref{pissi}) and $v(s)$ as in
Proposition \ref{pr3.1}$(c)$ with
\begin{equation}
\label{ez}
E_z= 2(v_1 - v_0) \Delta u + 2 \int_0^1 \left( \frac{1}{2} \langle \dot x, \dot x\rangle -
V_{\Delta}(x,s)\right) ds.\quad \cvd
\end{equation}
\smallskip

Even though we are always considering differentiable curves,
it is convenient now to introduce a wider space of curves which is useful
for a variational approach to (\ref{pissi}).
Fixed two points $x_0$, $x_1 \in \mo$, define the set
\[
\Omega^1(x_0,x_1) = \{x \in H^1(I,\mo) : x(0) = x_0 ,\; x(1) = x_1\},
\]
where $H^1(I,\mo)$ is the Sobolev space containing the absolutely
continuous curves from $I$ to $\mo$ with finite integral of $\langle\dot{x},\dot{x}\rangle$.
It is well-known (see, for example,  \cite[Proposition 2.2]{CFS}) that a curve
$x\in\Omega^{1}(x_{0},x_{1})$ is a classical solution of the problem (\ref{pissi})
if and only if it is a critical point of the functional
\begin{equation} \label{eJ}
{\cal J}_{\Delta}: x \in \Omega^1(x_0,x_1) \longmapsto
\ {1\over 2}\ \int_0^1 \langle\dot x,\dot x\rangle\ ds
\ - \int_0^1 V_{\Delta}(x,s)\ ds \in \R.
\end{equation}

Even if the existence of critical points for this functional is
the most classical problem in calculus of variations, its
complete solution for a natural case as the one we are
interested ($V_\Delta$ is differentiable and may behave quadratically at infinity)
has been obtained only very recently by the authors
by means of variational methods and Ljusternik--Schnirelmann theory
(see \cite{CFS}).
In particular, \cite[Theorem 1.1]{CFS} implies:

\begin{propo}
\label{mainpropo}
Let $(\mo,\langle\cdot,\cdot\rangle)$ be a
complete (connected) $n$--dimensional Riemannian manifold.
Let $V\in C^1(\m_{0} \times I,\R)$
be such that
\begin{equation}
\label{quadratic}
 V(x,s) \le \lambda d^2(x,\bar x) + \mu d^{p}(x,\bar x) + k  \quad
\mbox{for all $(x,s) \in \m_{0} \times I$,}
\end{equation}
for some $p\in [0,2[$, $\bar x \in \m_{0}$ and (positive) real numbers $\lambda, \mu, k$.

If $\lambda < \pi^2/2$, then for all $x_0, x_1 \in \mo$
there exists at least one solution of the corresponding
problem (\ref{pissi}) which is an absolute minimum of ${\cal J}_\Delta$.

Moreover, if $\mo$
is not contractible in itself, there exists a sequence of solutions $\{x_k \}_k$
such that ${\cal J}_\Delta (x_k) \rightarrow +\infty$ if $k \to +\infty$.
\end{propo}
Hence, the following result in PFWs can be stated:

\begin{theo}
\label{pisi}
Let $(\m,\langle\cdot,\cdot\rangle)$,  $\m = \mo \times \R^2$, be a PFW and
fix $u_0, u_1 \in \R^2$, with  $u_0 \leq u_1$. Suppose that:
\begin{enumerate}
\item[$(H_1)$] $(\mo,\langle\cdot,\cdot\rangle)$
is a  {\em complete} $n$--dimensional Riemannian manifold;

\item[$(H_2)$] there exist $p \in [0,2[ $, $\bar x \in \mo$ and (positive)
real numbers $R_0 $, $R_1, R_2$ such that
for all $(x,u) \in \mo \times [u_0,u_1]$ it is
\begin{equation} \label{etprinc}
H(x,u) \ge - (R_0 d^2(x,\bar x) + R_1 d^p(x,\bar x) + R_2) .
\end{equation}
\end{enumerate}

Then, two points $z_0=(x_0, v_0, u_0), z_1=(x_1, v_1, u_1) \in \m$
can be joined by a geodesic if
\begin{equation} \label{bound}
R_0 (u_1-u_0)^2 < \pi^2.
\end{equation}

Moreover, under these hypotheses, if $\mo$ is not contractible
in itself, then there exist infinitely many spacelike geodesics connecting $z_0$ and $z_1$.
\end{theo}

\proof
It is straightforward from Propositions \ref{prlema} and \ref{mainpropo}.
To check that, in the case $\mo$ non-contractible,  infinitely many
{\em spacelike} geodesics $\{z_k\}_k$ exist, recall that, using (\ref{eJ}),
the value of their energy $E_{z_k}=E_z(x_k)$ in (\ref{ez}) is
\[
E_z(x_k) = 2 \Delta v \Delta u + 2 {\cal J}_{\Delta}(x_k),
\]
where $\Delta v$ and $\Delta u$ are constant and $2 {\cal J}_{\Delta}(x_k)\rightarrow +\infty$
if $k \to +\infty$.
\cvd

\begin{rema}
\label{rema10} {\rm
(1) Inequality (\ref{etprinc}) does not impose any condition on the growth
of $H(x,u)$ with respect to $u$, because we assume that $u$ lies in the compact interval $[u_0, u_1]$.
That is, one can consider $R_0, R_1, R_2$ as continuous functions of $u$
and impose $(H_2)$ for $x\in \mo$ at each fixed $u\in [u_0,u_1]$.
Then, the bound (\ref{bound})  would be obtained just putting
$R_0 = \max \{R_0(u): u\in [u_0,u_1 ]\}$.

(2) Hypothesis $(H_2)$ can be deduced as a consequence of alternative
inequalities involving  either the $\mo$-  gradient $\nabla_x$ or
Hessian Hess$_x$ of $H$. In fact, $(\ref{etprinc})$ will hold if
one of the following two conditions is satisfied:
\begin{enumerate}
\item[$(i)$] $H\in C^1(\m_{0} \times [u_0,u_1],\R)$ and for all $(x,u) \in \mo \times [u_0,u_1]$ it is
\begin{equation} \label{egrad}
\langle \nabla_xH(x,u), \nabla_xH(x,u) \rangle^{1/2} \leq 2 R_0 d(x,\bar x) + R_1;
\end{equation}

\item[$(ii)$] $H\in C^2(\m_{0} \times [u_0,u_1],\R)$ and
there exists $K \geq 0$
such that for all $x \in \m_{0}$ with $d(x,\bar x) \ge K$ it is
\begin{equation}
\label{ehess}
{\rm Hess}_x H(x,u)[\xi,\xi] \le 2 R_0 \langle \xi,\xi\rangle
\qquad  \mbox{for all $\xi\in T_x\m_{0}, u\in [u_0,u_1]$}
\end{equation}
\end{enumerate}
(it is also possible  to replace each one of these inequalities by a $\limsup$,
as in \cite[Remark 1.2]{CFS}).

Remarkably, (\ref{egrad}) as well as (\ref{ehess})
imply also the different inequality for $H$ which was needed
for the result on completeness stated in Corollary \ref{co3.3}.}
\end{rema}
Notice that, if $(H_2)$ holds in each compact interval $[u_0,u_1]$ with $R_0=0$,
then the geodesic connectedness is obtained. More precisely:

\begin{coro} \label{corocon}
A PFW is
geodesically connected if $(H_1)$ in Theorem \ref{pisi} holds and
\begin{enumerate}
\item[$(H_2')$] there exist $\bar x \in \mo$, (positive) continuous functions $R_1(u), R_2(u)$
and $p(u) < 2$ such that for all $(x,u) \in \mo \times \R$ it is
 \begin{equation} \label{etprinc'}
H(x,u) \ge - (R_1(u) d^{p(u)}(x,\bar x) + R_2(u)).
\end{equation}
\end{enumerate}
\end{coro}
The following example (inspired by \cite[Example 3.6]{CFS}) proves that
condition (\ref{bound}) is the best estimate one can obtain if
$H$ grows quadratically with respect to $x$.

\begin{ex}
\label{esempio}
{\em Consider an (exact) PFW, $\m = \mo \times \R^2$, $\mo = \R^n$ with $H(x,u) = - |x|^2$.
Obviously, condition ($H_2$) is satisfied
($R_0 = 1$, $R_1=R_2 = 0$, $\bar x = 0$)
and Theorem \ref{pisi} ensures  geodesic connectedness for any pair
of points $z_0 = (x_0,v_0,u_0)$, $z_1 = (x_1,v_1,u_1)$ such that
$|u_1 - u_0| < \pi$.
On the contrary, there are non--geodesically connectable points
$z_0$, $z_1$ with $|\Delta u| = |u_1 - u_0| = \pi$.
In fact, recall that in this model it is $V_{\Delta}(x(s)) = \frac{\pi^2}{2} |x(s)|^2$
and the corresponding Riemannian problem (\ref{pissi}) becomes
\[
\left\{
\begin{array}{l}
\ddot x (s) + \pi^2\ x(s) = 0 \\
x(0) = x_0 ,\;\; x(1) = x_1.
\end{array}
\right.
\]
Clearly, taken $x_0=0$ and $x_1 \neq0$, this problem has no solution
and the corresponding two points in $\m$ are non--connectable.}
\end{ex}

\subsection{Connectedness by causal geodesics} \label{s5}

Now, in order to give
an Avez-Seifert type result, let us recall that
any PFW $\m$ is time oriented (see Section \ref{s2}) and, fixed any
$z_0=(x_0, v_0, u_0) \in \m$, its causal future is defined as
\begin{eqnarray*}
J^+(z_0) &=& \{z \in \m :\, \mbox{$z = z_0$ or there is a future directed}\\
&&\phantom{\{z \in \m :}\, \mbox{piecewise smooth causal curve in $\m$ from $z_0$ to $z$}\}.
\end{eqnarray*}
Thus, if $z_1=(x_1, v_1, u_1) \in J^+(z_0)$, then
 $u_0\leq u_1$ by  (\ref{udot}).

\begin{theo}
\label{prrr}
Let $(\m,\langle\cdot,\cdot\rangle)$ be a PFW, $\m = \mo \times \R^2$. Fix
$z_0=(x_0, v_0, u_0)$, $z_1=(x_1, v_1, u_1) \in \m$ with $z_1 \in J^+(z_0)$.

Assume that $(H_1)$, $(H_2)$ and (\ref{bound}) in Theorem \ref{pisi} hold.

Then, there exists a future directed causal geodesic from $z_0$ to $z_1$
which has maximum length among all the causal curves connecting these endpoints.
\end{theo}

\proof
Let $z : I \rightarrow \m$,
$z(s)=(x(s),v(s),u(s))$, be a future directed causal curve such that
$z(0) = z_0$, $z(1) = z_1$. We can assume $u_0<u_1$. Otherwise,
as $\dot u(s)  \geq 0$ (see (\ref{udot})), $u(s)$ would be constant
and $z(s)$ must be a lightlike pregeodesic of $\Pi_{u_0}$
(see Corollary \ref{corocoro}). We will also assume $\dot u(s)>0$ for all $s$ because,
otherwise, from standard arguments in Causality Theory a longer timelike curve with
the same endpoints could be found. Moreover, $z(s)$ will be considered reparametrized
in such a way that $\dot u(s)$ is constant, i.e., $s=(u(s)-u_0)/\Delta u$, $\Delta u=u_1-u_0$.

By Proposition \ref{mainpropo} and Theorem \ref{pisi},
$z_0$ and $z_1$ can be joined by a geodesic $\bar z : I \to {\cal M}$,
$\bar z(s) = (\bar x(s),\bar v(s),\bar u(s))$, such that
$\bar x$ is a minimum point of ${\cal J}_{\Delta}$ in $\Omega^1(x_0,x_1)$,
while $\bar u$ and $\bar v$ are defined as in Proposition \ref{pr3.1} $(a)$, $(c)$.
Our aim is to prove that
\be \label{etimelike}
E_{\bar z} \leq \int_0^1E_z(s) ds,
\ee
where $E_{\bar z}$ is the energy of the geodesic $\bar z$, as computed from (\ref{ez}), and
$E_z(s)= \langle \dot z(s), \dot z(s)\rangle (\leq 0)$ on $I$. In fact,
inequality (\ref{etimelike}) proves not only that $\bar z(s)$ is
causal but also that it has minimum energy among the considered curves
and, by a standard application of Cauchy-Schwarz inequality, maximum length
among all connecting causal curves.

Now, recall that ${\cal J}_{\Delta}(\bar x) \leq {\cal J}_{\Delta}(x)$ and,
using (\ref{eV}), (\ref{quarto}), (\ref{ez}) and (\ref{eJ}):
\begin{eqnarray*}
E_{\bar z} &\leq& 2(v_1 - v_0) \Delta u + 2 {\cal J}_\Delta(x) \\
&=& 2 \int_0^1 \left(\dot v \dot u
+ \frac{1}{2} \langle \dot x, \dot x\rangle
- V_{\Delta}(x, s)\right) ds =  \int_0^1E_z(s) ds,
\end{eqnarray*}
as required.
\cvd

As in the case of geodesic connectedness, our hypotheses for connection under causal curves
are sharp, as the following example shows.

\begin{ex}
\label{esempio1}
{\em Let $(\R^{n+2},\langle\cdot,\cdot\rangle)$ be the PFW
introduced in Example \ref{esempio}. We have already remarked
that $(H_1), (H_2)$ are satisfied but, for example,
the points $z_0 = (0,0,0)$ and $z_1 = (x_1,v_1,\pi), x_1\neq 0$ (which do not satisfy (\ref{bound}))
cannot be connected by a geodesic for any value of $v_1$. Anyway, $v_1$ can be chosen such that
$z_1 \in J^+(z_0)$. In fact, taken
$v_1 < 0$ with $|v_1|$ large enough, the connecting curve
$ z(s) = s \cdot (x_1, v_1, \pi)$, $s \in I$, is causal.}
\end{ex}

\begin{rema} \label{rema10bis}
{\em All the remarks to the hypotheses of Theorem \ref{pisi} still hold.
In particular, in the case that, instead of $(H_2)$,
the stronger assumption $(H'_2)$ in Corollary \ref{corocon} holds,
then inequality (\ref{bound}) will hold automatically; thus,
two points will be causally related if and only if they can be joined by
a causal (length--maximizing) geodesic.
In the forthcoming article \cite{FS} hypotheses
$(H_1)$ and $(H_2')$ are shown to imply global hyperbolicity, and, thus, in this case the result on existence of causal geodesics can be obtained as a consequence of classical Avez--Seifert's one. Nevertheless, our proof of Theorem \ref{prrr}
is based on completely different arguments,
and we obtain not only the result in the non--globally hyperbolic case $(H_2)$
but also some multiplicity results on timelike geodesics,
as we show next (see Theorem \ref{prrr2}
and Corollary \ref{prrr1}). On the other hand, the proof of global hyperbolicity
can be simplified by using Theorem \ref{prrr} (see \cite[Theorem 4.1, Remark 4.4]{FS}). }
\end{rema}
If the Riemannian part of a PFW is topologically non--trivial,
it has been proved the existence of infinitely many spacelike geodesics
(see Theorem \ref{pisi}). Now, under the same assumptions,
we are able to prove some multiplicity results for timelike geodesics.

\begin{theo}
\label{prrr2}
Let $\m = \mo \times \R^2$ be a PFW. Fix
$z_0=(x_0, v_0, u_0) \in \m$ and $(x_1, u_1) \in \m_{0}\times\R$ such that
$\Delta u = u_1 - u_0 \ne 0$.

Assume that $(H_1)$, $(H_2)$ and (\ref{bound}) in Theorem \ref{pisi} hold
and $\mo$ is not contractible in itself.

Then, either
\begin{eqnarray*}
\lim_{v\rightarrow -\infty} N(z_0, z_v) = +\infty&& \mbox{if $\Delta u > 0\quad $ or}\\
\lim_{v\rightarrow +\infty} N(z_0, z_v) = +\infty &&\mbox{if $\Delta u < 0$,}
\end{eqnarray*}
where $z_v = (x_1,v,u_1)$, $v \in \R$, and
$N(z_0,z_v)$ is the number of timelike geodesics from $z_0$ to $z_v$.
\end{theo}

\proof
In order to prove the multiplicity result, let us recall
some more details about
the variational and topological tools which are needed in the proof of
Proposition \ref{mainpropo}.
In fact, if $\mo$ is not contractible in itself,
by a Fadell and Husseini's result (cf. \cite{FH})
it follows that the manifold of curves $\Omega^1(x_0,x_1)$ has
infinite Ljusternik--Schnirelman category and contains
compact subsets of arbitrarily high category. Whence, since the
classical Ljusternik--Schnirelmann Theorem applies to
${\cal J}_{\Delta}$ in the given assumptions
(see \cite{CFS}), such a functional has infinitely many
critical points $\{x_k\}_k$ such that
\be
\label{critical}
{\cal J}_{\Delta}(x_k) = \inf_{A \in \Gamma_k} \sup_{x \in A}{\cal J}_{\Delta}(x)
 \qquad \mbox{for each $k \in \N$, $k \ge 1$,}
\ee
with
\[
\Gamma_k = \{ A \subseteq \Omega^1(x_0,x_1) :\; \cat_{\Omega^1(x_0,x_1)}(A) \ge k\}
\]
(here, $\cat_{\Omega^1(x_0,x_1)}(A)$ is the Ljusternik--Schnirelmann category
of $A$ with respect to $\Omega^1(x_0,x_1)$, i.e., the least number of
closed and contractible subsets of $\Omega^1(x_0,x_1)$ covering $A$; for more details
see, e.g., \cite{Pa1} or also \cite[Section 2.6]{Ma}).
Obviously, by the definition (\ref{critical}) it follows that
\[
{\cal J}_{\Delta}(x_1) \le {\cal J}_{\Delta}(x_2)\le \dots \le {\cal J}_{\Delta}(x_k)\le \dots\ .
\]
On the other hand, fixed $t \in \R$ and $z_t = (x_1,t,u_1)$, Proposition \ref{pr3.1}
and (\ref{ez}), (\ref{eJ}) imply that for any $k \ge 1$ the curve
$z_t^k(s) = (x_k(s),v_t^k(s),u_0+s \Delta u)$, $s \in I$, is a geodesic in $\m$
with energy
\be
\label{stimeenergy}
E_t^k = 2 (t - v_0) \Delta u + 2 J_\Delta (x_k)
\ee
if $v_t^k(s)$ is as in Proposition \ref{pr3.1}$(c)$.

Now, let $m \in \N$, $m \ge 1$, be fixed.
Whence, by (\ref{stimeenergy}), if $\Delta u > 0$ there exists $t_m < v_0$ such that
for all $t \le t_m$ the corresponding
$z_t^1,z_t^2, \dots,z_t^m$
are $m$ timelike geodesics joining $z_0$ to $z_t$;
while, on the other hand, if $\Delta u < 0$ the same result holds but choosing
$t_m > v_0$ and $t \ge t_m$.
\cvd

If  we replace condition $(H_2)$ in Theorem \ref{prrr2} by the stronger one $(H'_2)$, then
the further assumption (\ref{bound}) holds automatically since it is $R_0 = 0$
(in this case, the functional ${\cal J}_{\Delta}$ is bounded
from below for every $\Delta u = u_1 - u_0$);
thus, not only the same results
of Theorem \ref{prrr2} still hold but the same arguments
of its proof allow to prove a second multiplicity
estimate. More precisely, it can be proved that:

\begin{coro}
\label{prrr1}
Let $\m = \mo \times \R^2$ be a PFW. Fix
$z_0=(x_0, v_0, u_0)\in \m$ and $(x_1, v_1) \in \m_{0}\times\R$.

Assume that $(H_1)$ and $(H'_2)$ in Corollary \ref{corocon} hold
and let $\mo$ be not contractible in itself.

Then, either
\begin{eqnarray*}
\lim_{u\rightarrow -\infty} N(z_0, z_u) = +\infty&& \mbox{if $v_1 > v_0\quad$ or}\\
\lim_{u\rightarrow +\infty} N(z_0, z_u) = +\infty &&\mbox{if $v_1 < v_0$,}
\end{eqnarray*}
where $z_u = (x_1,v_1,u)$, $u \in \R$, and
$N(z_0,z_u)$ is the number of timelike geodesics from $z_0$ to $z_u$.
\end{coro}

\subsection{Application to exact gravitational waves} \label{s6}

At last, we want to apply the previous results to the
classical models of exact gravitational waves.
To this aim,
the previous Theorems \ref{pisi}, \ref{prrr}, \ref{prrr2},
Corollary \ref{prrr1} and Remarks \ref{rema10}(1), \ref{rema10bis}
can be summarized as follows:

\begin{coro} \label{coro0}
Let $(\m,\langle\cdot,\cdot\rangle)$, $\m = \mo \times \R^2$,
be a PFW such that $(H_1)$ holds and fix $\bar x \in \mo$.
Let $R_0(u), R_1(u), R_2(u)$ ($\ge 0$),
$p(u)<2$ be continuous functions and put, for any $u_0, u_1 \in \R$
(not necessarily $u_{0}\leq u_{1}$):
\[
R_0[u_0, u_1] = {\rm Max}\{R_0(u): u \in [u_0, u_1] \cup [u_1,u_0]\}
\]
(with $[u_i,u_j] = \emptyset$ if $u_i > u_j$, $i,j \in\{0,1\}$).
Then,
\begin{enumerate}
\item[(1)] if
$H(x,u) \ge - \left(R_1(u) d^{p(u)}(x,\bar x) + R_2(u)\right)$
for all $(x,u) \in \mo \times \R$,
then $\m$ is
geodesically connected;

\item[(2)] if
$H(x,u) \ge - \left(R_0(u) d^2(x,\bar x) + R_1(u) d^{p(u)}(x,\bar x) + R_2(u)\right)$
for all $(x,u)$  $ \in \mo \times \R$,
two points $z_0 = (x_0,v_0,u_0)$, $z_1 = (x_1,v_1,u_1) \in \m$
can be surely connected by means of a geodesic if
\begin{equation} \label{boundbis}
R_0[u_0,u_1] (u_1-u_0)^2 < \pi^2.
\end{equation}
\end{enumerate}
Moreover, when either the case (1) or the case (2) holds:
\begin{enumerate}
\item[$(a)$] if $z_1 \in J^+(z_0)$ there exists a length-maximizing causal geodesic connecting $z_0$ and $z_1$;

\item[$(b)$] if $\mo$ is not contractible in itself:
\begin{enumerate}
\item[$(i)$] there exist infinitely many spacelike geodesics
connecting $z_0$ and $z_1$,

\item[$(ii)$] the number of timelike geodesics from $z_0$ to $z_v=(x_1,v,u_1)$
goes to infinity when $v\rightarrow -\infty$ if it is $u_1 > u_0$
or when $v\rightarrow +\infty$ if it is $u_1 < u_0$.
\end{enumerate}
\end{enumerate}

Furthermore, only in the case (1), if $\mo$ is not contractible in
itself the number of timelike geodesics from $z_0$ to
$z_u=(x_1,v_1,u)$ goes to infinity when $u\rightarrow -\infty$ if
it is $v_1 > v_0$ or when $u\rightarrow +\infty$ if it is $v_1 <
v_0$.
\end{coro}
In particular, Corollary \ref{coro0}
is appliable to exact gravitational waves as follows:

\begin{propo}
\label{c1}
Let $(\R^4,ds^2)$ be an exact gravitational plane wave. Then,
the case (2) of Corollary \ref{coro0} holds with
\[
R_0[u_0, u_1] = {\rm Max}\{ (f^2 + g^2)^{1/2}(u): u \in [u_0, u_1] \cup [u_1,u_0]\}
\quad\mbox{and $R_1, R_2 \equiv 0$.}
\]
\end{propo}

\proof
Recall that, for an exact gravitational wave,
$H(\cdot, u)$ is a quadratic form with eigenvalues
$\pm (f^2 + g^2)^{1/2}(u)$. Thus, for all $x=(x_1, x_2)$, we have
$H(x,u) \geq -(f^2 + g^2)^{1/2}(u) |x|^2$ (the equality holds for the corresponding eigenvectors).
\cvd

\begin{rema} \label{rema1}
{\em Fixed $u_0$, the function $R_0[u_1]:= R_0[u_0,u_1]$ cannot decrease when $|u_1-u_0|$ grows.
Thus, on any exact gravitational wave
the left hand side of (\ref{boundbis}) must reach the value $\pi^2$
for some values of $u_1$, i.e., we can find unique $u_1^-, u_1^+ \in \R$ such that:
\begin{enumerate}
\item[$(i)$] $u_1^- < u_0 < u_1^+$ and

\item[$(ii)$] $R_0[u_1^{-}] = \frac{\pi^2}{(u_1^{-}-u_0)^2}, \quad \quad
R_0[u_1^{+}] = \frac{\pi^2}{(u_1^{+}-u_0)^2}$.
\end{enumerate}
Thus, {\em Corollary \ref{coro0} is applicable whenever $u \in\ ]u_1^-,u_1^+[$}. In particular, this yields a bound for the appareance of the first astigmatic conjugate point (see
\cite[pp. 486]{BEE}).
}\end{rema}
Even more, using the same idea of previous examples,
we can check that our hypotheses are the sharpest ones, even for exact polarized sandwich waves.

\begin{ex}
\label{esempio2}
{\em Let $(\R^4,d s^2)$ be an exact gravitational wave
such that in (\ref{gravitational}) it is
$f(u)=1$ on $[0,\pi]$, $f(u)=0$ out of
a compact subset and $g(u) \equiv 0$. Choose the points $z_0 = (0,0,0,0)$ and
$z_1 = (x_1,0, v_1,\pi)$ with $x_1 \ne 0$.
The same arguments in Examples \ref{esempio}, \ref{esempio1}
show that $z_0$ and $z_1$ cannot be connected by a geodesic
for any $v_1 \in \R$, even in the case that $-v_1>0$ is
large enough such to imply $z_1 \in J^+(z_0)$.
}\end{ex}


\end{document}